\documentclass[twocolumn,showpacs,preprintnumbers,amsmath,amssymb]{revtex4}

\usepackage{graphicx}
\usepackage{bm}

\usepackage{color}

\begin{document}

\title{Plateau border bulimia transition:\\
discontinuities expected in three simple experiments on 2D liquid foams}

\author{Pierre Rognon}
\author{Fran\c{c}ois Molino}
\author{Cyprien Gay}

\email{cyprien.gay@univ-paris-diderot.fr}

\affiliation{%
Centre de Recherche Paul Pascal, CNRS, UPR~8641,
Universit\'e de Bordeaux~1,
115 Av. Schweitzer, F--33600 PESSAC, France\\
Department of Endocrinology,
Institute of Functional Genomics, CNRS, UMR~5203,
INSERM U661, Universities of Montpellier~1 and~2,
141 rue de la Cardonille, F--34094 MONTPELLIER cedex 05, France\\
Mati\`{e}re et Syst\`{e}mes Complexes, 
Universit\'{e} Paris Diderot--Paris 7, CNRS, UMR~7057, 
B\^atiment Condorcet, Case courrier 7056, F--75205 PARIS cedex 13, France
}%

\date{\today}

\begin{abstract}
We describe the geometry of foams
squeezed between two solid plates (2D GG foams)
in two main asymptotic regimes:
fully dry floor tiles and dry pancakes.
We predict an abrupt transition between both regimes,
with a substantial change in the Plateau border radius.
This should be observable in different types of experiments
on such 2D GG foams:
when foam is being progressively dried or wetted,
when it is being squeezed further or stretched,
when it coarsens through film breakage or Oswald ripening.
\end{abstract}

\maketitle

\newcommand{\hide}[1]{}

\newcommand{\hs}{\hspace{0.6cm}}
\newcommand{\be}{\begin{equation}}
\newcommand{\ee}{\end{equation}}
\newcommand{\bee}{\begin{eqnarray}}
\newcommand{\eee}{\end{eqnarray}}

\newcommand{\mysection}{\section}

\newcommand{\transp}[1]{{#1}^T}
\newcommand{\trace}{{\rm tr}}

\newcommand{\si}{\sigma}
\newcommand{\unittensor}{I}
\newcommand{\surfacetension}{\gamma}
\newcommand{\specificsurface}{\Sigma}
\newcommand{\siinterf}{\si^{\rm interf}}
\newcommand{\philiq}{\phi}

\newcommand{\dimratio}{k}
\newcommand{\truebidi}{E-F}

\newcommand{\pgas}{p_{\rm g}}
\newcommand{\pliq}{p_{\rm l}}
\newcommand{\Dp}{\Delta p}
\newcommand{\pwall}{p_{\rm wall}}
\newcommand{\pdis}{\Pi_{\rm d}}
\newcommand{\piosm}{\pi_{\rm osm}}
\newcommand{\dilatancy}{\chi}

\newcommand{\Sbub}{S} 
\newcommand{\Abub}{{\cal A}} 
\newcommand{\Atot}{{\cal A}_{\rm tot}} 
\newcommand{\Vbub}{\Omega} 
\newcommand{\Vliq}{\Omega_{\rm liq}} 
\newcommand{\Vpb}{V_{\rm Pb}} 
\newcommand{\sppb}{S_{\rm hPb}} 
\newcommand{\Pbradius}{R} 
\newcommand{\pPbradius}{\Pbradius_{\rm ps}}  

\newcommand{\Vwn}{V_{\rm wn}} 
\newcommand{\shpb}{\sppb} 
\newcommand{\svpb}{S_{\rm vPb}} 

\newcommand{\va}{\vec{a}}
\newcommand{\vb}{\vec{b}}
\newcommand{\vc}{\vec{c}}
\newcommand{\vu}{\vec{u}}
\newcommand{\vv}{\vec{v}}
\newcommand{\vw}{\vec{w}}

\mysection{Introduction}\label{Sec:Intro}

As a physical system, liquid foams display 
many interesting properties~\cite{weaire_hutzler_1999_book,weaire2007rf,hohler2005rlf}.
Among these, 2D foams attract a widespread 
interest : their local structure is
more easily observable than that of three-dimensional foams,
and more amenable to theoretical calculations.

In the present work, we present a comprehensive
(though approximate) geometric description
of 2D foams squeezed between two solid plates (``GG'' foams).
A thorough examination of these geometric properties suggests 
non-trivial consequences which should be observable 
through simple experimental protocols.

This approach is extended to variational properties like stresses and 
dilatancy in a separate work~\cite{dilatancy_geometry_letter_2008,dilatancy_geometry_article_2008}.

In Section~\ref{Sec:2d_gg_foam_vademecum},
we will go through an original and detailed geometrical description
of a 2D GG foam.
In Section~\ref{Sec:discontinuities},
we will highlight, as a surprising consequence, the existence of abrupt variations
of the Plateau border radius with tunable parameters of the foam,
and describe three different experimental situations
where such variations should be observable.

\mysection{2D glass-glass foam geometrical Vademecum}\label{Sec:2d_gg_foam_vademecum}

In the present section,
we shall provide a geometrical description
and some corresponding results
for {\em two-dimensional foams} squeezed between two glass plates.
More specifically, the calculations are conducted
for {\em ordered}, {\em monodisperse} foams.
Of course, if some local disorder is present,
the results should not be affected tremendously.
But the foam should nevertheless be {\em spatially homogeneous}:
there should be no large-scale gradient 
in cell thickness, bubble volume, etc.
Indeed, such a gradient would generate
subtle effects related to the osmotic pressure and force balances,
which would feed the discussion substantially.

In the remaining of this article,
we call $P$ the typical perimeter of the bubble,
defined as the perimeter of the rounded polygon
that constitutes the bubble, as seen from above,
and $\Pbradius$ the corresponding radius of curvature
of the Plateau border (see Fig.~\ref{Fig:pancake}, left).
We call $H$ the distance between both solid plates,
and $\pPbradius$ the radius of curvature 
of the pseudo Plateau borders 
(see Fig.~\ref{Fig:pancake}, bottom right).
We also call $\Vbub$ the bubble volume,
and $\Vliq$ the volume of liquid per bubble.
In the monodisperse case, the liquid volume fraction $\philiq$ thus verifies:
\be
\label{Eq:Omega_liq}
\philiq=\frac{\Vliq}{\Vbub+\Vliq},
\hs{i.e.}\hs
\Vliq=\Vbub\frac{\philiq}{1-\philiq}.
\ee
We assume that $\Vbub$ remains constant:
we target experimental conditions under which the applied stresses are not sufficient
to compress the gas phase significantly. 
This is the case at atmospheric pressure,
unless the bubble size is on the order of a micron or smaller.

\subsection{Geometry: floor tile versus pancake regime}

\begin{figure}
\begin{center}
\resizebox{1.0\columnwidth}{!}{%
\includegraphics{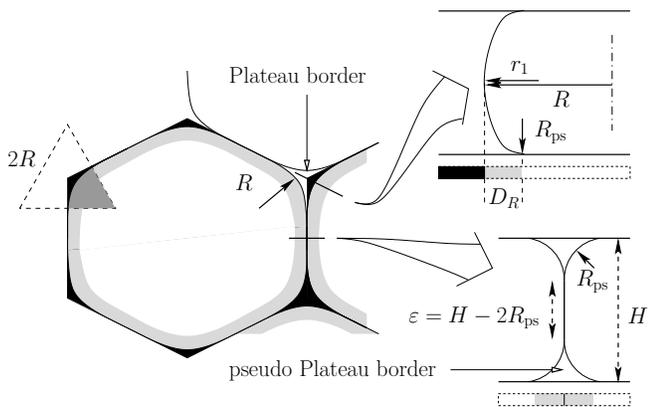}
}
\end{center}
\caption{Pancake conformation of a bubble
squeezed between two solid plates (distance $H$): 
approximate geometry.
Left: top view.
The variable $P$ denotes 
the average perimeter of the bubbles in such a top view
(outer perimeter of the light grey region).
The variable $\Vbub$ is the volume of the bubble gas
(full thickness of the white region
and part of the thickness of the light grey region),
and $\Abub$ is defined as $\Abub=\Vbub/H$.
The variable $\Atot$ is the total (gas and liquid)
projected surface area per bubble (white and light grey and black regions).
Right: two different (magnified) cross-sections (side views)
with matching greyscale.
As seen from above, the contact between two bubbles,
for an ordered monodisperse foam,
is typically along a straight line (left).
In this region, the {\em pseudo Plateau border} (see bottom right drawing) has a uniform curvature
of radius $\pPbradius$.
By contrast, in the {\em genuine Plateau border} region (left drawing), 
the section of the gas-liquid interface
is approximately elliptical in shape (top right drawing), 
with principal radii of curvature $r_1$ and  $\Pbradius$ at mid-height,
and $\pPbradius$ and zero at the plates.
The width $D_R$ of the curved region
is intermediate between $\pPbradius$ and $H/2$
while $r_1$ is larger than $H/2$.
}
\label{Fig:pancake}
\end{figure}

\begin{figure}
\begin{center}
\resizebox{0.8\columnwidth}{!}{%
\includegraphics{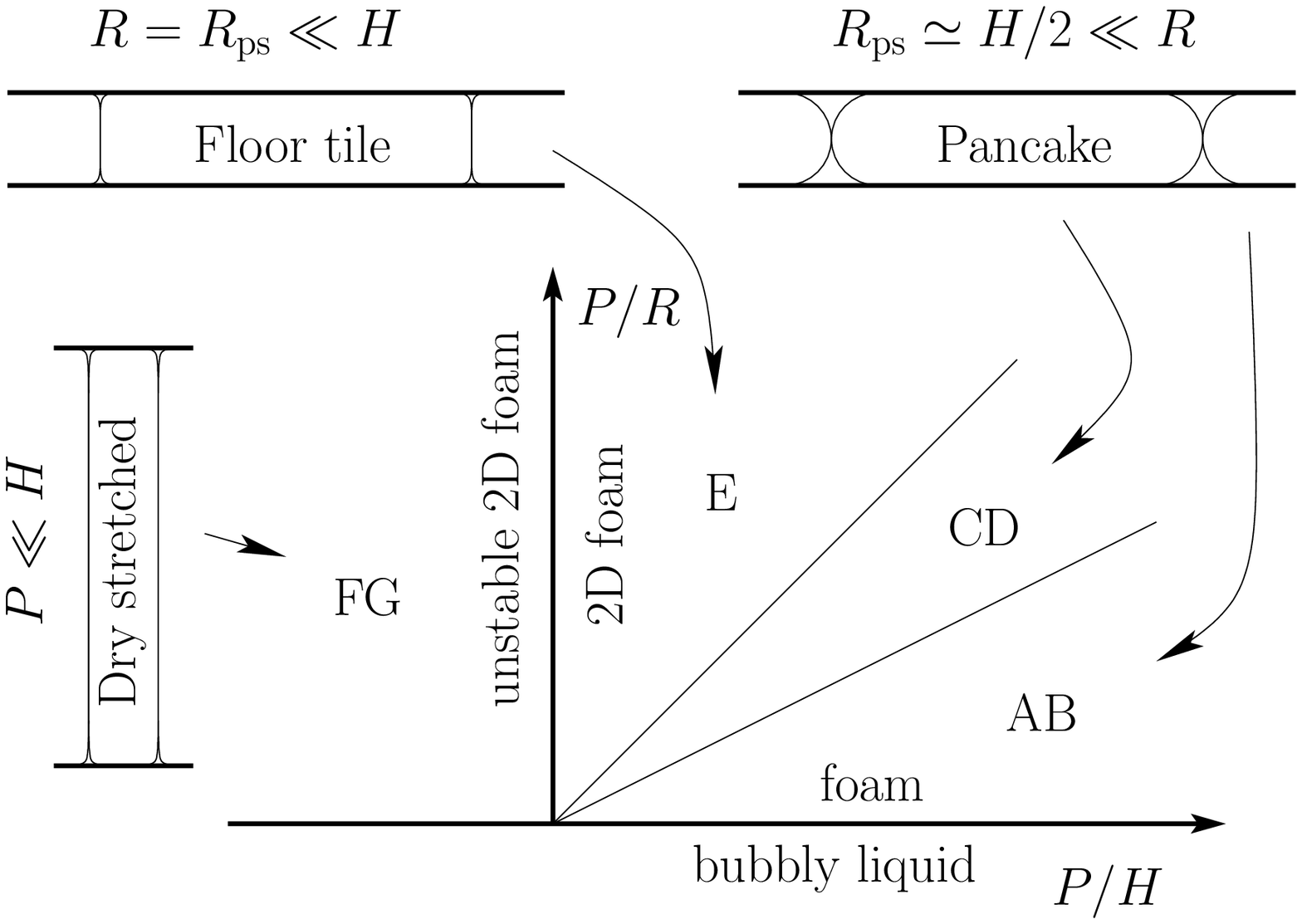}
}
\end{center}
\caption{Log-log representation of the regimes of a glass-glass 2D foam 
with low liquid fraction ($\philiq\ll 1$),
in terms of the bubble perimeter $P$,
the Plateau border radius $\Pbradius$
and the cell height $H$.\\
The bubble perimeter $P$ is measured at mid-height of the cell:
it is the outer perimeter of the light grey ribbon
in Fig.~\ref{Fig:pancake}.
Such a foam can be found in two main configurations.
In the floor tile situation (regimes E and FG)
the pseudo Plateau borders
are much thinner than the cell height
($\Pbradius=\pPbradius\ll H$).
By contrast, in the pancake regime (AB and CD),
although the overall liquid volume fraction $\philiq$
is still much smaller than 1,
facing pseudo Plateau borders almost join
($H-2\pPbradius\ll H\ll\Pbradius$),
and each bubble has a pancake-like shape.
More precisely, 
it is useful to subdivide the pancake situation
into regimes AB and CD defined
by Table~\ref{tab:geometrical_transitions}:
in regime AB (resp. CD),
the liquid content of the Plateau border
(resp. pseudo Plateau border) dominates.
The corresponding expressions 
for the liquid volume fraction and for several geometrical quantities
are indicated in Table~\ref{tab:geom_gg}.
Note that regime FG, which starts with $P\approx H$,
corresponds to a non-realistic 2D GG foam,
where the height is larger than the perimeter.
The limit of a truly two-dimensional foam
is obtained with $H\rightarrow\infty$.
Note also that when $P=2\pi\,R$ (horizontal axis),
the bubbles become independent in the liquid.}
\label{Fig:diagram_pancake_regimes_ABCD}
\end{figure}

The main contributions to the quantity of liquid per bubble
are pictured on Fig.~\ref{Fig:pancake}
and can be calculated from simple geometrical arguments:
\be
\label{Eq:Vliq_PR}
\Vliq \simeq
(2-\pi/2)\,P\,\pPbradius^2
+(2\sqrt{3}-\pi)\,\Pbradius^2 H
\ee
The first term corresponds 
to the volume of the pseudo Plateau borders.
As seen from above,
they correspond to the light grey regions
in Fig.~\ref{Fig:pancake}.
Each portion of their interfaces
has the shape of a quarter of a circular cylinder
(see Fig.~\ref{Fig:pancake}, bottom right).
The second term in Eq.~(\ref{Eq:Vliq_PR})
corresponds to the genuine Plateau borders
(black regions in Fig.~\ref{Fig:pancake}).

Eq.~(\ref{Eq:Vliq_PR}) indicates
that the squeezed 2D foam lies between two asymptotic regimes
depending on volume fraction and geometry.
They are pictured on Fig.~\ref{Fig:diagram_pancake_regimes_ABCD}.

\begin{enumerate}
\item When the Plateau border radius $\Pbradius$ is much larger
than the sample thickness $H$ 
(regimes AB and CD of Fig.~\ref{Fig:diagram_pancake_regimes_ABCD}),
each bubble takes the form of a thick ``pancake'',
and its edge is like a half cylinder with radius $H/2$,
which is the largest accessible value
for the radius of the pseudo Plateau borders
(see paragraph~\ref{Sec:elliptical_pancake} below
for a finer description).
\item In the reverse limit, the bubbles
are shaped more like ``floor tile'', with sharp edges
(regime E of Fig.~\ref{Fig:diagram_pancake_regimes_ABCD}):
this time, the Plateau borders are like fine threads
pinned on both solid plates,
and each pseudo Plateau borders resembles
a stretched, fine thread, glued on one of the solid plate
and joining the attachment points of two Plateau borders;
in this case, pseudo and genuine Plateau border radii coincide.
\end{enumerate}
On Fig.~\ref{Fig:diagram_pancake_regimes_ABCD},
we have also pictured regime FG,
which does not correspond to realistic 2D GG foams:
it is useful to obtain the limit of ideal 2D foams,
for which the solid plates are so far apart
that the volume of the pseudo Plateau borders
can be entirely neglected.
This regime, although unrealistic,
bridges the gap between 2D GG foam considered here
and ideal 2D foams.
This is useful in particular for discussing static
dilatancy~\cite{weaire_2003_2747,rioual_2005_117,dilatancy_geometry_letter_2008,dilatancy_geometry_article_2008}.

In the pancake regimes (AB and CD),
one has $\pPbradius\simeq H/2$ and hence:
\be
\label{Eq:Vliq_PR_ABCD}
\Vliq^{\rm ABCD} \simeq (1/2-\pi/8)\,P\,H^2
+(2\sqrt{3}-\pi)\,\Pbradius^2 H
\ee
In regime AB, the Plateau border contribution dominates
(black regions on Fig.~\ref{Fig:pancake}).
Hence:
\be
\Vliq^{\rm AB} \simeq (2\sqrt{3}-\pi)\,\Pbradius^2 H
\ee
Conversely, in regime CD,
the pseudo Plateau borders (light grey regions
on the top-view in Fig.~\ref{Fig:pancake})
contain most of the liquid:
\be
\Vliq^{\rm CD} \simeq (1/2-\pi/8)\,P\,H^2
\ee

In regimes E and FG ($\Pbradius\ll H$, floor tile regime), 
the radius of curvature $\pPbradius$
of the pseudo Plateau borders 
is equal to that of the Plateau borders:
\be
\label{Eq:pPbradius_EFG}
\pPbradius^{\rm EFG}=\Pbradius
\ee
and Eq.~(\ref{Eq:Vliq_PR}) reduces to:
\be
\label{Eq:Vliq_PR_EFG}
\Vliq^{\rm EFG} \simeq
[(2-\pi/2)\,P+(2\sqrt{3}-\pi)\,H]\,\Pbradius^2
\ee
Far away from regime FG ($P\gg H \gg \Pbradius$), one has:
\be
\Vliq^{\rm E} \simeq (2-\pi/2)\,P\,\Pbradius^2
\ee

The above distinction between the pancake regime
and the floor tile regime,
based solely on the overall shape of the bubbles
(rounded edges {\em versus} sharp edges),
is sufficient to provide 
the results of Eqs.~(\ref{Eq:Vliq_PR_ABCD})
and~(\ref{Eq:Vliq_PR_EFG})
concerning the liquid volume per bubble
in the various regimes (see Table~\ref{tab:geom_gg}).

But for other purposes, in particular concerning 
dilatancy~\cite{dilatancy_geometry_letter_2008},
one must refine the geometrical description
of the bubbles in such a GG foam in the pancake regime.
That is the scope of the remaining of the present Section.

\begin{table}
\begin{tabular}{ | c | l | }
\hline
Regime & Conditions
\\\hline
AB
& \begin{tabular}{ c | c }
$\Pbradius<P$ &
$\Pbradius^2 > P\,H$
\\
($\philiq<\philiq_c$)
& (dominant Plateau borders)
\end{tabular}
\\\hline

CD
& \begin{tabular}{ c | c }
$\Pbradius^2 < P\,H$ &
$\Pbradius > H$
\\
(dominant pseudo-Plateau borders)
& (pancake)
\end{tabular}
\\\hline

E
& \begin{tabular}{ c | c }
$\Pbradius < H$
&
$P > H$
\\
(floor tile)
& 
(2D)
\end{tabular}
\\\hline

FG
& \begin{tabular}{ c | c }
$\Pbradius < H$
&
$P > \Pbradius$
\\
(stretched 2D)
& 
($\philiq<\philiq_c$)
\end{tabular}
\\\hline
\end{tabular}

\caption{Four regimes for a 2D glass-glass foam.
Regimes AB and CD correspond to pancake-shaped bubbles,
while regime E corresponds to a foam made of floor tile shaped bubbles.
In regimes CD and E, most of the liquid
is located in the pseudo Plateau borders,
while in regime AB, 
the Plateau borders themselves have a greater volume.
Regime FG does not correspond to stable 2D GG foams
(they would destabilize into 3D foams),
but the limit $H\rightarrow\infty$
corresponds to an ideal, truly two-dimensional foam.}
\label{tab:geometrical_transitions}
\end{table}

\subsection{Pancake regime with elliptical cross-section}
\label{Sec:elliptical_pancake}

Only numerical tools such as Surface Evolver~\cite{Brakke92}
can provide an exact description of the bubbles
when a more thorough geometrical description
than the above approximation is needed.

In the present paragraph devoted to the pancake regime,
we consider an intermediate approximation
based on an elliptical description
of the pseudo Plateau border cross-section
and derive various geometrical properties.
A similar derivation (with a simpler geometrical description)
was carried out recently~\cite{marchalot_2008}
in the context of coarsening experiments.

The shape of an inter-bubble film is a surface
whose total curvature is related to the pressure difference
between both bubbles:
\be
\frac{1}{\rho_1}+\frac{1}{\rho_2}=\frac{p_B-p_A}{2\surfacetension}
\ee
where $\rho_1$ and $\rho_2$ are the principal radii of curvature,
$p_A$ and $p_B$ are the gas pressure in bubbles A and B,
and where the surface tension of the film
is assumed to be equal to twice that of a single interface, $\surfacetension$.
In the present work, for simplicity we assume that all bubbles
have equal pressures, so that the films are planar:
$\rho_1=\rho_2=\infty$.

Since the films are assumed planar,
the pseudo Plateau borders are circular in cross-section
(radius $\pPbradius$, see Fig.~\ref{Fig:pancake} bottom right)
and the total curvature of the gas-liquid interface
is therefore $1/\pPbradius$.

In the Plateau border region, 
the meniscus is roughly toroidal in shape:
the Plateau border is axisymmetric (radius $\Pbradius$),
and the interface is curved more strongly in the vertical direction
(with a radius of curvature close to $H/2$).
Let $r(z)$ be the equation for its surface,
in cylindrical coordinates as measured from its axis of symmetry.
The constant total curvature condition reads:
\be
\label{Eq:total_curvature_r_z}
\frac{-r^{\prime\prime}}{(1+r^{\prime 2})^{3/2}}
+\frac{1}{r\,(1+r^{\prime 2})^{1/2}}
={\rm const}
=\frac{1}{\pPbradius}
\ee
where $r^{\prime}={\rm d}r/{\rm d}z$
and $r^{\prime\prime}={\rm d}^2r/{\rm d}z^2$.
The first term is the curvature within a meridian plane,
while the second term is the curvature
within a plane perpendicular to the meridian curve.
The value of the constant is chosen
so as to match the total curvature
found for the same gas-liquid interface 
in the pseudo Plateau borders,
namely $1/\pPbradius$.

Of course, Eq.~(\ref{Eq:total_curvature_r_z}) 
for the surface shape can be solved numerically.
Here, we shall simply take the limit $\Pbradius\gg H$
(which corresponds to asymptotic regimes AB and CD),
where bubbles are not pressed very strongly
against one another.
This has two consequences:
{\em (i)} the interbubble films are very small in height
($H-2\pPbradius\ll H$)
and {\em (ii)} the shape of the interface in the Plateau border
can be approximated as elliptical in cross-section.

With this elliptical approximation
(see Fig.~\ref{Fig:pancake} top right),
the various geometrical parameters
can be obtained very easily.
At mid-height, the azimuthal radius of curvature
is simply the Plateau border radius $\Pbradius$.
Since the total curvature is known to be $1/\pPbradius$,
the radius of curvature $r_1$ in the meridian plane at mid-height
can be obtained through:
\be
\frac{1}{r_1}+\frac{1}{\Pbradius}
=\frac{1}{\pPbradius}
\ee
Since the length of the greater semi-axis 
of the ellipse is simply $H/2$, one has:
\be
(H/2)^3=r_1^2\,\pPbradius
\ee
With these two equations
and keeping in mind that $\Pbradius\gg H$,
one can then deduce the various lengths
that appear on Fig.~\ref{Fig:pancake} (top right),
including the smaller semi-axis $D_R$
given by $D_R^3=\pPbradius^2\,r_1$:
\bee
r_1 &\simeq& \frac{H}{2}\,\left[1+\frac16\frac{H}{\Pbradius}\right] \label{Eq:r_1_ellipse}\\
\pPbradius^{\rm ABCD} &\simeq& \frac{H}{2}\,\left[1-\frac13\frac{H}{\Pbradius}\right] \label{Eq:pPbradius_ellipse}\\
D_R &\simeq& \frac{H}{2}\,\left[1-\frac16\frac{H}{\Pbradius}\right] \label{Eq:D_R_ellipse}
\eee

Note that the current description in terms
of Plateau borders with an elliptical cross-section
and pseudo Plateau borders with an circular cross-section
is not entirely consistent:
the crossover between both regions
implies that the interface adopt
an intermediate shape near the cuspy corners of the Plateau border.
In the present limit $\Pbradius\gg H$,
this discrepancy can be safely neglected.
Similarly, the elliptical approximation
for the solution of Eq.~(\ref{Eq:total_curvature_r_z})
is sufficient.
Nevertheless, all subsequent results 
will not be expressed beyond the first order in $H/\Pbradius$.

The height $\varepsilon$ of the interbubble film
is given by:
\be
\varepsilon=H-2\pPbradius \simeq \frac13\,\frac{H^2}{\Pbradius} 
\label{Eq:epsilon_ellipse}
\ee
Note that with a stronger geometrical assumption
(circular rather than elliptical cross-section
for the torus in the Plateau border),
Marchalot {\em et al.}~\cite{marchalot_2008}
obtain almost the same result.
Only the numerical prefactor differs
($1/2$ in Ref.~\cite{marchalot_2008}
instead of $1/3$ in Eq.~(\ref{Eq:epsilon_ellipse}) above).

\subsection{Liquid volume per pancake}
\label{Sec:liquid_volume_per_pancake}

Let us now refine Eq.~(\ref{Eq:Vliq_PR})
with the elliptical approximation presented above.

The first term, $(2-\pi/2)\,P\,\pPbradius^2$,
corresponds to the pseudo Plateau borders
(light grey regions of Fig.~\ref{Fig:pancake}
when seen from above)
and already assumes that their cross-section is circular.
Hence, this part of the equation remains correct
in the refined description,
provided the value of $\pPbradius$
is taken from Eq.~(\ref{Eq:pPbradius_ellipse})
and provided we subtract the part of the perimeter
that corresponds to the Plateau borders:
\be
\label{Eq:pBvolume_ellipse}
\frac{4-\pi}{8}\,H^2\,(P-2\pi\,\Pbradius)
\,\left[1-\frac{2\,H}{3\Pbradius}\right]
\ee
where $2\pi\,\Pbradius$
is the sum of the six Plateau border perimeters
of a typical bubble
({\em i.e.}, the total length of the separation
between light grey and black regions 
in Fig.~\ref{Fig:pancake}).

\begin{figure}
\begin{center}
\resizebox{0.8\columnwidth}{!}{%
\includegraphics{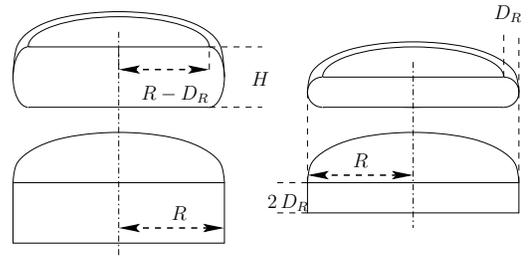}
}
\end{center}
\caption{Contribution to the liquid in a Plateau border.
Within a Plateau border, the gas-liquid interface
has roughly the shape of the outer surface of an elliptical torus (top left).
Hence, part of the liquid in the Plateau borders of a bubble
consists in the region between such an elliptical torus 
and the circular wall of a vertical cylinder (bottom left)
with the same outer radius $\Pbradius$.
To calculate the volume of this contribution more easily,
we reduce the height of both the torus (top right)
and the cylinder (bottom right) by such a factor 
that the torus has a circular cross-section.
}
\label{Fig:elliptical_torus}
\end{figure}

The second term,
$(2\sqrt{3}-\pi)\,\Pbradius^2 H$,
still correctly represents the part of the Plateau borders
that spans the entire gap $H$ between both plates
(black regions on Fig.~\ref{Fig:pancake}).
We must add, however,
the volume of liquid located above and below 
the elliptical region of the interface
(which corresponds to the curved part 
of the grey ribbons that surround
the black Plateau border core regions).
If we put together all these Plateau border edges for one single bubble,
we obtain a fully axisymmetric region of thickness $H$
that is described on Fig.~\ref{Fig:elliptical_torus} (left).
Its volume can be deduced from that of a rounded flat cylinder
with thickness $2\,D_R$ and with a toroidal edge
(see Fig.~\ref{Fig:elliptical_torus} right).
The volume of such a rounded flat cylinder (top right) can be expressed as:
\be
2\,D_R\,\pi\,(\Pbradius-D_R)^2
+2\pi\,(\Pbradius-D_R)\,\frac{\pi\,D_R^2}{2}
+\frac{4\pi\,D_R^3}{3}
\ee
where the first term is the volume of a flat cylinder 
with radius $\Pbradius-D_R$ and height $2\,D_R$,
the second term is that of a semi-cylinder
whose radius is $D_R$ and whose length is the perimeter
of the first flat cylinder.
Finally, the third term is the correction
that accounts for the curvature of the semi-cylinder,
and it is equal to the volume of a sphere of radius $D_R$.
Subtracting these terms from the volume
$2D_R\,\pi\Pbradius^2$ of the outer flat cylinder
(see Fig.~\ref{Fig:elliptical_torus} bottom right)
and multiplying this result by a factor $H/(2D_R)$,
we obtain the desired liquid contribution:
\be
\frac{H}{2D_R}\left[
\pi(4-\pi)\,D_R^2\,\Pbradius
-\frac{\pi(10-3\pi)}{3}\,D_R^3
\right]
\ee

Combining this with Eq.~(\ref{Eq:pBvolume_ellipse})
and with the second term of Eq.~(\ref{Eq:Vliq_PR}),
we finally obtain:
\bee
\Vliq\simeq 
\frac{4-\pi}{8}\,H^2P
+(2\sqrt{3}-\pi)\,\Pbradius^2H
-\frac{4-\pi}{12}\,\frac{H^3P}{\Pbradius}&&\ \nonumber\\
+\frac{\pi}{12}\,H^3+\frac{\pi(10-3\pi)}{72}\,\frac{H^4}{\Pbradius}&&
\label{Eq:Vliq_PR_ABCD_refined_complete}
\eee
The relative magnitude of the first three terms
changes between regimes AB and CD 
(see Fig.~\ref{Fig:diagram_pancake_regimes_ABCD}),
while the fourth and fifth terms are always negligible
in both regimes:
\be
\begin{array}{ l r }
{\rm regime\,AB}: & T_2 \gg T_1 \gg T_3 \gg T_4 \gg T_5 \\
{\rm regime\,CD}: & T_1 \gg (T_2\,{\rm and}\,T_3) \gg T_4 \gg T_5
\end{array}
\ee
Because the fourth and fifth terms
are always negligible,
we shall retain only the first three terms
in Eq.~(\ref{Eq:Vliq_PR_ABCD_refined_complete}):
\be
\label{Eq:Vliq_PR_ABCD_refined}
\Vliq^{\rm ABCD}\simeq 
\frac{4-\pi}{8}\,H^2P
\,\left[1-\frac{2H}{3\Pbradius}\right]
+(2\sqrt{3}-\pi)\,\Pbradius^2H
\ee
which refines the result of Eq.~(\ref{Eq:Vliq_PR_ABCD}).
In other words, we use the elliptical approximation
to obtain the value $\pPbradius$ of the pseudo Plateau border radius,
see Eq.~(\ref{Eq:pPbradius_ellipse}).
But we then calculate the volume as if the shape of the interface
in the Plateau borders were identical to that in the
pseudo Plateau borders, {\em i.e.},
with circular (radius $\pPbradius$) rather than elliptical sections 
(see Fig.~\ref{Fig:pancake} bottom right).
This is legitimate because in the limit 
of small liquid fractions ($\philiq\ll 1$),
the total length of the Plateau borders ($2\pi\,\Pbradius$)
is much smaller than that of the pseuso Plateau borders ($P$).

\subsection{Liquid volume fraction}
\label{Sec:liquid_volume_fraction}

From Eqs.~(\ref{Eq:Vliq_PR_EFG})
and~(\ref{Eq:Vliq_PR_ABCD_refined}),
we derive the liquid volume fraction 
$\philiq=\Vliq/(\Atot\,H)$ in the foam,
both in the pancake regime and in the floor tile regime:
\bee
\label{Eq:philiq_ABCD}
\philiq^{\rm ABCD}&\simeq& 
\frac{4-\pi}{8}\,\frac{H\,P}{\Atot}
\,\left[1-\frac{2H}{3\Pbradius}\right]
\nonumber\\
&&+(2\sqrt{3}-\pi)\,\frac{\Pbradius^2}{\Atot}\\
\label{Eq:philiq_EFG}
\philiq^{\rm EFG}&\simeq&
\frac{4-\pi}{2}\,\frac{\Pbradius^2\,P}{\Atot\,H}
+(2\sqrt{3}-\pi)\,\frac{\Pbradius^2}{\Atot}
\eee
The corresponding asymptotic values of $\philiq$ in all sub-regimes
are indicated in Table~\ref{tab:geom_gg}.
It is useful to express the liquid fraction
in terms of the volume $\Vbub$ of the bubble itself.
Using $\philiq/(1-\philiq)=\Atot\,H\,\philiq/\Vbub$
and the total volume of the bubble and liquid 
$\Atot\,H=\Vbub+\Vliq$,
the above equations become:
\bee
\label{Eq:philiq_ABCD_Vbub}
\frac{\philiq^{\rm ABCD}}{1-\philiq^{\rm ABCD}}&\simeq&
\frac{4-\pi}{8}\,\frac{H^2\,P}{\Vbub}
\,\left[1-\frac{2H}{3\Pbradius}\right]\nonumber\\
&&+(2\sqrt{3}-\pi)\,\frac{\Pbradius^2\,H}{\Vbub}\\
\label{Eq:philiq_EFG_Vbub}
\frac{\philiq^{\rm EFG}}{1-\philiq^{\rm EFG}}&\simeq& 
\frac{4-\pi}{2}\,\frac{\Pbradius^2\,P}{\Vbub}
+(2\sqrt{3}-\pi)\,\frac{\Pbradius^2\,H}{\Vbub}
\eee
%

\subsection{Specific surface area}
\label{Sec:surface_area_per_bubble}

With the same approximation as above,
let us calculate the total surface area of a bubble.
As it is directly related to the foam energy,
it is useful when deriving the elastic modulus 
of the foam~\cite{dilatancy_geometry_article_2008}.

The contribution from the top and the bottom of a bubble
to the in-plane component is the white region 
of the top view in Fig.~\ref{Fig:pancake}:
$$
2\,\left\{\Atot-[2\sqrt{3}\,\Pbradius^2-\pi\,(\Pbradius-D_R)^2]
-(P-2\pi\,\Pbradius)\,\pPbradius\right\}
$$
In this expression, the term $-2\sqrt{3}\,\Pbradius^2$
corresponds to one third of the dashed-line triangle 
with edge length $2\Pbradius$ in Fig.~\ref{Fig:pancake}
and similarly for the other Plateau borders of the bubble.
The third term, $\pi\,(\Pbradius-D_R)^2$,
corresponds to the dark grey sector in the triangle.
The next term, $-(P-2\pi\,\Pbradius)\,\pPbradius$,
corresponds to the wall surface area 
that does not touch the bubble
along the pseudo Plateau borders (light grey ribbon).

Making the approximation $D_R\simeq\pPbradius$
in the Plateau border region 
as mentioned at the end of Paragraph~\ref{Sec:liquid_volume_per_pancake}
and using $\pPbradius\ll P$, this becomes:
\be
\label{Eq:surface_area_top_and_bottom}
2\,\Atot-(4\sqrt{3}-2\pi)\,\Pbradius^2
-2\,P\,\pPbradius
\ee

The vertical films contribution is:
\be
\label{Eq:surface_area_films}
P\,(H-2\pPbradius)
\ee

The contribution from the menisci,
considered as circular quarter cylinders with radius $\pPbradius$,
can be written as:
\be
\label{Eq:surface_area_menisci}
2P\,\int_0^{\frac{\pi}{2}}\pPbradius\,{\rm d}\theta
=\pi\,P\,\pPbradius
\ee

Hence, the specific surface area $\specificsurface$ of the foam,
equal to the bubble total surface area
divided by the total volume $\Atot\,H$,
includes all three contributions above:
\bee
\Atot\,H\,\specificsurface&\simeq&
2\,\Atot-(4\sqrt{3}-2\pi)\,\Pbradius^2
-2\,P\,\pPbradius\nonumber\\
&&+P\,(H-2\pPbradius)
+\pi\,P\,\pPbradius
\nonumber\\
\specificsurface&\simeq&\frac{2}{H}
-(4\sqrt{3}-2\pi)\,\frac{\Pbradius^2}{\Atot\,H}\nonumber\\
&&+\frac{P}{\Atot}
-(4-\pi)\,\frac{P\,\pPbradius}{\Atot\,H}
\eee
The final result is then derived
both in the pancake regime where $\pPbradius$
is given by Eq.~(\ref{Eq:pPbradius_ellipse})
and where $\pPbradius\ll\Pbradius$ 
(regimes AB and CD of Fig.~\ref{Fig:diagram_pancake_regimes_ABCD})
and in the floor tile regime where $\pPbradius=\Pbradius$ (regimes E and FG):
\bee
\label{Eq:specific_surface_area_ABCD}
\specificsurface_{\rm ABCD}&\simeq&\frac{2}{H}
-(4\sqrt{3}-2\pi)\,\frac{\Pbradius^2}{\Atot\,H}\nonumber\\
&&+\frac{\pi-2}{2}\,\frac{P}{\Atot}
+\frac{4-\pi}{6}\,\frac{P\,H}{\Atot\,\Pbradius}\\
\label{Eq:specific_surface_area_EFG}
\specificsurface_{\rm EFG}&\simeq&\frac{2}{H}
+\frac{P}{\Atot}
-(4-\pi)\,\frac{P\,\Pbradius}{\Atot\,H}\nonumber\\
&&-(4\sqrt{3}-2\pi)\,\frac{\Pbradius^2}{\Atot\,H}
\eee

These results are reproduced in Table~\ref{tab:geom_gg}.

\begin{table}
\begin{tabular}{ | c c | c | c | }
\hline
Quantity & Eqs. & Value & Regimes \\
\hline
pseudo Plateau & $\pPbradius$
 & $\frac{H}{2}\,\left(1-\frac{H}{3\Pbradius}\right)$ & ABCD \\
border radius & (\ref{Eq:pPbradius_EFG},\,\ref{Eq:pPbradius_ellipse})
 & $\Pbradius$ & EFG \\
\hline
volume & & $(2\sqrt{3}-\pi)\,\Pbradius^2\,H$ & ABFG \\
of liquid & $\Vliq$
 & $\frac{4-\pi}{8}\,P\,H^2$ & CD \\
per bubble & (\ref{Eq:Vliq_PR_ABCD},\,\ref{Eq:Vliq_PR_EFG})
 & $\frac{4-\pi}{2}\,P\,\Pbradius^2$ & E \\
\hline
liquid & & $(2\sqrt{3}-\pi)\,\frac{\Pbradius^2}{\Atot}$ & ABFG \\
volume fraction & $\philiq$ 
& $\frac{4-\pi}{8}\,\frac{P\,H}{\Atot}$ & CD \\
($\philiq=\frac{\Vliq}{\Atot\,H}$) 
& (\ref{Eq:philiq_ABCD}-\ref{Eq:philiq_EFG_Vbub})
 & $\frac{4-\pi}{2}\,\frac{\Pbradius^2\,P}{\Atot\,H}$ & E \\
\hline
& & $\frac{2}{H}-\frac{(4\sqrt{3}-2\pi)\,\Pbradius^2}{\Atot\,H}$ & AB \\
specific & $\specificsurface$ & $\frac{2}{H}+\frac{\pi-2}{2}\,\frac{P}{\Atot}$ & CD \\
surface area & (\ref{Eq:specific_surface_area_ABCD}-\ref{Eq:specific_surface_area_EFG})
& $\frac{2}{H}+\frac{P}{\Atot}$ & E \\
& & $\frac{P}{\Atot}$ & FG \\
\hline
\end{tabular} 
\caption{Geometrical properties
of a two-dimensional glass-glass foam.
The numbers refer to the relevant series of equations
and the letters to the regimes
of Fig.~\ref{Fig:diagram_pancake_regimes_ABCD}:
pancake regime (AB and CD) and floor tile regime (E and FG).
In each expression, the terms are ordered by decreasing magnitudes.
}
\label{tab:geom_gg}
\end{table}

\mysection{Predicted non-trivial discontinuities
in three experimental conditions:
the bulimic Plateau borders}\label{Sec:discontinuities}

We shall now analyse the geometrical quantities 
derived in the previous section
and show that they predict non-trivial discontinuities.

When preparing an undeformed 2D GG foam,
the experimental conditions determine 
the value of three independent variables,
for instance the liquid fraction $\philiq$,
the bubble volume $\Vbub$ and the cell thickness $H$.

Once these quantities are fixed, 
the bubble perimeter $P$ depends principally
on $\Vbub$ and $H$, with the scaling
\be
\label{Eq:perimeter_scaling}
\Vbub\simeq P^2\,H,
\ee
since $P^2$ scales like the bubble surface area $\Atot$.
It also depends more weakly 
on the degree of disorder of the foam
and on the bubble size distribution.
Finally, it is also sensitive to the foam deformation,
and this may be at the origin of static dilatancy, 
as we show elsewhere~\cite{dilatancy_geometry_letter_2008}.

The Plateau border radius $\Pbradius$
also results from $\philiq$, $\Vbub$ and $H$
and from the foam equilibrium.

Then other quantities, such as those derived 
in Section~\ref{Sec:2d_gg_foam_vademecum} above,
are also determined.

\subsection{Description and origin of the transition}\label{Sec:transition_and_origin}

The above calculations have a striking consequence:
an intrinsic instability of regime CD
leads to discontinuities in the dimension of the Plateau borders
when varying continuously $\philiq$, $H$ or $P$.

In Section~\ref{Sec:2d_gg_foam_vademecum}, for simplicity,
the liquid fraction $\philiq$
was expressed in terms of $P$, $H$ and $\Vbub$ or $\Atot$.
It appears that in regime CD,
$\philiq$ does not depend on $\Pbradius$ (see Table~\ref{tab:geom_gg}).
A more refined analysis, based on Eq.~(\ref{Eq:philiq_ABCD_Vbub}),
reveals that there is a slight dependence
of $\philiq$ on $\Pbradius$:
\be
\frac{8}{4-\pi}\,\frac{\Vbub\,\philiq}{H^2\,P} - 1
\simeq \frac{8(2\sqrt{3}-\pi)}{4-\pi}\,\frac{\Pbradius^2}{H\,P}
-\frac23\,\frac{H}{R}
\ee
which shows that $\philiq\simeq\frac{4-\pi}{8}\,\frac{H^2\,P}{\Vbub}$
changes only by a factor close to unity
when $\Pbradius$ changes from small values comparable to $H$
to much larger values comparable to $\sqrt{H\,P}$.
In other words, the Plateau border radius $\Pbradius$
depends very strongly on the other parameters in this region.
When $H^{4/3}\,P^{2/3}\ll\Pbradius^2\ll H\,P$,
Eq.~(\ref{Eq:philiq_ABCD_Vbub}) yields:
\be
\label{Eq:Pbradius_C}
\Pbradius\simeq\sqrt{\frac{\Vbub}{(2\sqrt{3}-\pi)\,H}\,
\left(\philiq-\frac{4-\pi}{8}\,\frac{H^2\,P}{\Vbub}\right)}
\ee
Similarly, when $H \ll P \ll H^{2/3}\,P^{1/3}$:
\be
\label{Eq:Pbradius_D}
\Pbradius\simeq\frac{\frac23 H}
{1-\frac{8}{4-\pi}\,\frac{\philiq\,\Vbub}{H^2\,P}}
\ee

Thus, abrupt transitions can thus occur 
for the Plateau border radius $\Pbradius$
without any significant changes in other characteristic foam parameters.

This effect can be described in physical terms as follows.
When the foam goes across regime CD,
the fact that the Plateau border radius $\Pbradius$
is the only variable that varies significantly,
indicates that some liquid is exchanged between the Plateau borders
and the pseudo Plateau borders.
But as can be seen from Table~\ref{tab:geometrical_transitions},
in regime CD the volume of the Plateau borders
is small compared to that of the pseudo Plateau borders.
Hence, this exchange of liquid 
is of limited relative importance for the foam,
which explains why other variables are affected only marginally.

\subsection{Three experimental situations}\label{Sec:instability_three_experiments}

\begin{figure}
\begin{center}
\resizebox{1.0\columnwidth}{!}{%
\includegraphics{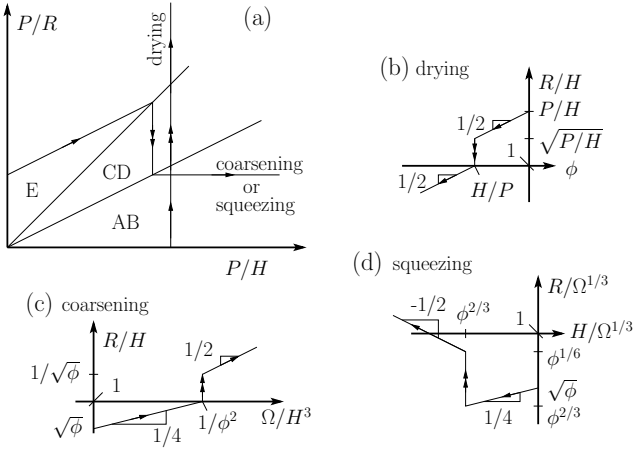}
}
\end{center}
\caption{Structural evolution of a 2D GG foam
undergoing three different experiments.
The rapid evolution through regime CD is symbolized by double arrows.
{\em (a)} Upon drying, the foam goes from regime AB to regime E
(in the notations of  Fig.~\ref{Fig:diagram_pancake_regimes_ABCD}).
By contrast, when it is squeezed or when it coarsens spontaneously,
it evolves from regime E to regime AB.
{\em (b-d)} Schematic Plateau border radius evolution
during three types of experiments.
{\em (b)} Drying: constant $P$ and $H$, decreasing $\philiq$.
{\em (c)} Coarsening: constant $\philiq$ and $H$, increasing $\Vbub$ and $P$.
{\em (d)} Squeezing: constant $\philiq$ and $\Omega$, 
decreasing $H$ and increasing $P$.
}
\label{Fig:three_experiments}
\end{figure}

Let us now imagine three experimental situations 
where the present ``bulimia'' effect should be apparent
in a GG-foam. The foam should be subjected to:
{\em (1)} progressive drying,
{\em (2)} coarsening (through film rupture or Oswald ripening),
{\em (3)} progressive squeezing.
Again, let us insist on the fact that the present predictions
rely on the assumption that the foam is almost at equilibrium
and that there is no spatial inhomogeneity in the foam.

\subsubsection{Drying foam}\label{Sec:drying}

Let us consider a GG-foam being progressively dried
(decreasing $\philiq$),
with constant bubble thickness $H$ and volume $\Vbub$,
hence almost constant perimeter $P$.

Starting near the liquid-solid transition,
with liquid fraction $\philiq$ comparable to unity,
the foam evolves vertically upwards in the diagram
of Fig.~\ref{Fig:three_experiments}a,
at first in regime AB, then in regime CD
(very rapidly as a result of the strong variation 
of $\Pbradius$ mentioned in Section~\ref{Sec:transition_and_origin}),
and finally in regime E.

The variations of the Plateau border radius $\Pbradius$
are plotted on Fig.~\ref{Fig:three_experiments}b
as a function of the liquid fraction $\philiq$.
Initially, it is comparable to the bubble perimeter,
it decreases as $\sqrt{\philiq}$ in regime AB,
it then jumps down through regime CD
to reach roughly $H/2$,
then it again decreases as $\sqrt{\philiq}$ in regime E:
\bee
\Pbradius_{\philiq\simeq 1} &\simeq& P\\
\Pbradius_{\rm AB} &\simeq& P\,\sqrt{\philiq}\\
H &\lesssim& \Pbradius_{\rm CD} \lesssim \sqrt{P\,H}\\
\Pbradius_{\rm E} &\simeq& \sqrt{P\,H}\,\sqrt{\philiq}
\eee
Of course, Eq.~(\ref{Eq:perimeter_scaling}) must be used
if the thickness $H$ and the bubble volume $\Vbub$
are prefered to the perimeter $P$ as constant parameters.

\subsubsection{Coarsening foam}\label{Sec:coarsening}

Let us now consider a GG-foam that coarsens progressively
(increasing typical bubble volume $\Vbub$ and perimeter $P$),
with constant thickness $H$ and liquid fraction $\philiq$.
This coarsening can result from various 
phenomena~\cite{weaire_hutzler_1999_book}.
When inter-bubble film are not very stable,
neighbouring bubbles may coalesce
and the average bubble size increases.
Alternatively, coarsening can result 
from continuous Oswald ripening:
gas diffusion between neighbouring bubbles
leads to a net flux from small bubbles to large bubbles
as a result of the larger pressure in the smaller bubbles;
large bubbles grow even larger, small bubbles vanish
and the size distribution evolves to larger length scales.

Starting near the 2D stability transition,
with a horizontal size (or perimeter $P$)
comparable to the thickness $H$,
the foam evolves as shown on the diagram
of Fig.~\ref{Fig:three_experiments}a,
at first in regime E, then in regime CD
(very rapidly as a result of the strong variation 
of $\Pbradius$ mentioned in Section~\ref{Sec:transition_and_origin}),
and finally horizontally in regime AB.

The variations of the Plateau border radius $\Pbradius$
are plotted on Fig.~\ref{Fig:three_experiments}c
as a function of the bubble volume $\Vbub$.
It increases like $\Vbub^{1/4}$ in regime E,
it then jumps up through regime CD
from $H/2$ to roughly $H/\sqrt{\philiq}$,
then it again increases like $\sqrt{\Vbub}$ in regime AB:
\bee
\Pbradius_{\rm E} &\simeq& (\Vbub\,H)^{1/4}\,\sqrt{\philiq}\\
H &\lesssim& \Pbradius_{\rm CD} \lesssim H/\sqrt{\philiq}\\
\Pbradius_{\rm AB} &\simeq& \sqrt{\Vbub\,\philiq/H}
\eee
Of course, Eq.~(\ref{Eq:perimeter_scaling}) must be used
if the perimeter $P$ is prefered to the bubble volume $\Vbub$
as the main variable.

In coarsening experiments such as those conducted 
by Marchalot {\em et al.}~\cite{marchalot_2008},
a (delicate) measurement of the Plateau border radius $\Pbradius$
will be necessary to observe the phenomenon described above clearly.

\subsubsection{Squeezing a foam}\label{Sec:squeezing}

Let us now consider a GG-foam that is being squeezed progressively
(decreasing thickness $H$)
at constant bubble volume $\Vbub$
and liquid fraction $\philiq$.

Starting near the 2D stability transition,
with a horizontal size (or perimeter $P$)
comparable to the thickness $H$,
the foam evolves as shown on the diagram
of Fig.~\ref{Fig:three_experiments}a,
at first in regime E, then in regime CD
(very rapidly as a result of the strong variation 
of $\Pbradius$ mentioned in Section~\ref{Sec:transition_and_origin}),
and finally horizontally in regime AB.

The variations of the Plateau border radius $\Pbradius$
are plotted on Fig.~\ref{Fig:three_experiments}d
as a function of the sample thickness $H$.
It is non-monotonic:
it decreases like $H^{1/4}$ in regime E,
it then jumps up through regime CD,
then it increases like $1/\sqrt{H}$ in regime AB:
\bee
\Pbradius_{\rm E} &\simeq& (\Vbub\,H)^{1/4}\,\sqrt{\philiq}\\
\Vbub^{1/3}\,\philiq^{2/3} &\lesssim& \Pbradius_{\rm CD} 
\lesssim \Vbub^{1/3}\,\philiq^{1/6}\\
\Pbradius_{\rm AB} &\simeq& \sqrt{\Vbub\,\philiq/H}
\eee
Of course, Eq.~(\ref{Eq:perimeter_scaling}) must be used
if the perimeter $P$ is prefered to the thickness $H$
as the main variable.

\mysection{Conclusion}

By examining the geometry of 2D ``GG''-foams
(located between two parallel solid plates),
we found that these foams should display
a rather strong ``bulimia'' transition
where Plateau borders swallow
(going from retime E to regime AB)
a large amount of liquid
when the foam parameters are varied in a narrow region.
This effect, which is specific to 2D ``GG''-foams, 
should be observable in at least three types of experiments,
for which we provided first-order predictions.
Of course, there may be other experimental situations.
Let us insist again on the fact that the present predictions
are restricted to foams almost at equilibrium
and with no spatial inhomogeneity.
This ``bulimia'' effect should affect
other phenomena, such as dilatancy; this is discussed 
elsewhere~\cite{dilatancy_geometry_letter_2008,dilatancy_geometry_article_2008}.

\mysection*{Acknowledgements}

We gratefully acknowledge fruitful discussions
with participants of the GDR 2983 Mousses (CNRS),
in particular with Julien Marchalot, 
Marie-Caroline Jullien and Isabelle Cantat
about the Plateau border geometry.
This work was supported by the Agence Nationale de la Recherche (ANR05).

\appendix

\bibliography{bulimia}
\end{document}